\documentclass[12pt,a4paper,reqno]{amsart}
\usepackage{amsmath}
\usepackage{amsfonts}
\usepackage{amssymb}
\usepackage{graphicx,psfrag}                    
\usepackage{listings}                           
\usepackage{algorithm}
\usepackage{algorithmic}
\numberwithin{equation}{section}

\theoremstyle{plain}

\theoremstyle{definition}

\newtheorem{definition}[subsection]{Definition}

\renewcommand{\leq}{\leqslant}

\newsavebox{\proofbox}
\savebox{\proofbox}{\begin{picture}(7,7)%
  \put(0,0){\framebox(7,7){}}\end{picture}}

\def\emph#1{{\it #1}}
\def\textbf#1{{\bf #1}}

\begin{document}

\title{A New Algorithm for the Subtraction Games}

\author{Guanglei He}
\address{Sichuan University jinjiang college,CHINA
}
\email{guangleihe@gmail.com}

\author{Zhihui Qin}

\email{zhihuiqindoris@gmail.com}

\begin{abstract}
Subtraction games is a class of combinatorial games. It was solved since the \emph{Sprague-Grundy Theory} was put forward. This paper described a new algorithm for subtraction games. The new algorithm can find win or lost positions in subtraction games. In addition, it is much simpler than Sprague-Grundy Theory in one pile of the games.
\end{abstract}


\maketitle

\section{Introduction}

Subtraction games is a class of impartial combinatorial games. The subtraction games is satisfies the following conditions.
\begin{itemize}
\item There are two-player game involved a pile of chips.
\item a finite set $S$ of positive integers called the \emph{subtraction set}. Subtraction sets $S = {s_1, s_2,... , s_k}$will be ordered $s_1 < s_2 < ・ ・ ・ < s_k$.
\item The two players move alternately, subtracting some $s$ chips such that $s \in S$.
\item The game ends when a position is reached from which no moves are possible for the player whose turn it is to move. The last player to move wins.
\end{itemize}

Since the \emph{Sprague-Grundy Theory} was proposed by Roland P.Sprague (in 1935) and independently Patrick M.Grundy (in 1939)(\cite{RJN}),people realized how to judge the game's position which is a $P$-position or an $N$-position (in \cite{EZ1} and \cite{EJR}), and how to move it to make sure win by mathematical methods.

However, the theory is so special that make us difficult to understand how to solve the problem. This paper described a new algorithm to solve subtraction games, and the algorithm make us simple to understand what happened. It used  $P$-position to find $N$-position, so we named it PTFN algorithm.

\section{Some Definitions}
   Before presenting our results, we need to recall some notations and terminologies.

In section 3, this paper need through the definitions of $P$-position and $N$-position to prove the correctness of the PTFN algorithm. $P$-positions' and $N$-positions' definition from the \cite{TF1}. It first noted by Ernst Zermelo \cite{EZ1} in 1912. $P$-positions and $N$-positions are defined recursively by the following three statements.
\begin{definition}$P$-position and $N$-position
\begin{itemize}
\item All \emph{terminal positions} are $P$-positions.
\item From every $N$-position, there is at least one move to a $P$-position.
\item From every $P$-position, every move is to an $N$-position.
\end{itemize}
\end{definition}
The \emph{terminal position} means the game has no chip to move.

\textbf{Sprague-Grundy theory} use a very special way to find all positions of the subtraction games. Sprague-Grundy theory was represented by \emph{Sprague-Grundy function} :$G(x)$,as $G(x) = 0$ it means the point $x$ is $P$-position ; as $G(x) \neq{} 0$ ,the point $x$ is $N$-position. Sprague-Grundy theory is famous, you can realize that in many papers such as \cite{TF1},\cite{EJR} etc, so this paper doesn't repeat it.

The following example is constructed for to describe how to use Sprague-Grundy function, this paper used the example comparison new algorithm to Sprague-Grundy theory.

\subsubsection*{Example 1} :As an example, the subtraction games with subtraction set $S = \{1, 3, 7 ,8\}$, Let us analyze it starts with a pile of n chips has a representation as a game (in the example it will let n = 21). Here $X = \{0, 1, . . . , n\}$ is the set of vertices. The empty pile is terminal, so $F(0)$, the empty set.  Also there are $F(1) = \{0\}, F(2) = \{1\},F(3) = \{1,0\}$, and for $2 \leq{} k \leq{} n, F(k) = \{k-8, k-7, k-3,k-1\}$ etc ,Figure.1.

\begin{figure}[htb]
\centering
\includegraphics[width=3.8in]{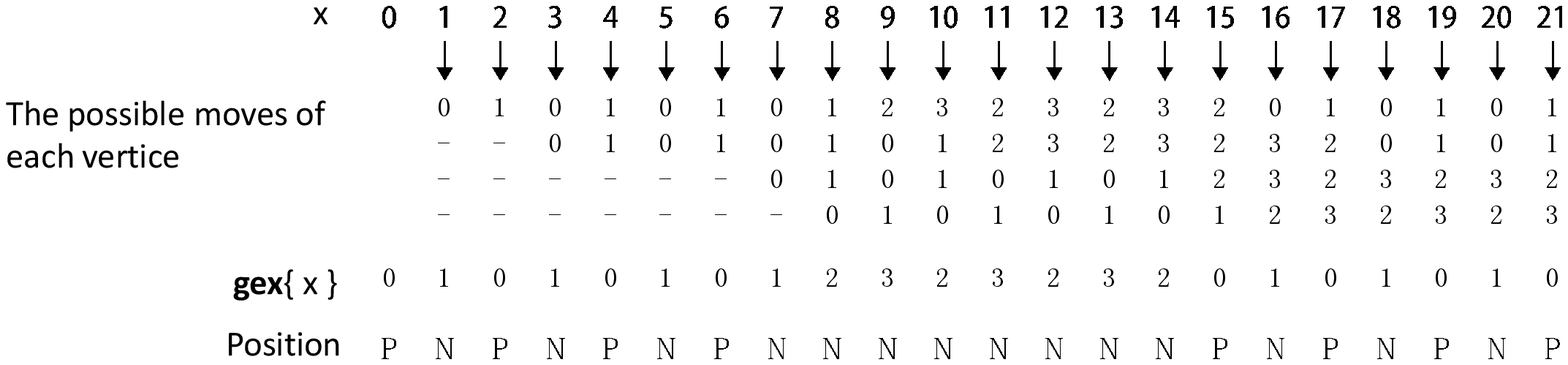}
\caption{The Subtraction Game with $S = \{1,3,7,8\}$}
\end{figure}

\section{PTFN algorithm}
For the subtraction games, this paper put forward a new algorithm to find the game's position. It is similar to Sprague-Grundy function ,but it simpler and more efficient than that. And it was named  PTFN($P$-position To Find $N$-position) algorithm. Just as its name , the algorithm used $P$-position to find $N$-position. It is easy to find all the game's positions.

At first, this paper just only considers one pile of chips in the subtraction games.

Specifically, it will use a array $X[n + 1]$ as a pile of $n$ chips game. Let $S$ be a set of positive integers. A probably move consists of removing $s$ chips from the pile where $s \in{} S$. Last player to move wins.

As this paper analyzed the subtraction games with subtraction set $S = \{1,3,7,9\}$. There is exactly a terminal position, namely $X[0]$. Then $X[0 + 1],X[0 + 3],X[0 + 7],X[0 + 8]$ must be $N$-position. they can be moved to $X[0]$. Then $X[2]$ is a $P$-position , it legal move to $X[2 + 1],X[2 + 3],X[2 + 7],X[2 + 8]$ , so these points must be $N$-positions. Just like that. It wouldn't finish this work until all positions of elements were found.
Now , it used pseudocode to describe the PTFN algorithm (Algorithm 1).
\begin{algorithm}
\centering
\caption{PTFN for one pile}
\label{alg1}
\begin{algorithmic}[1]
\STATE $SIZE\_OF\_PILE \leftarrow{} n$ \hfill \COMMENT{\textit{$n$ is size of pile}}
\STATE $SIZE\_OF\_SET \leftarrow{} s$  \hfill \COMMENT{\textit{$s$ is size of the set s}}
\newline
\STATE $X[SIZE\_OF\_PILE + 1]$             \hfill \COMMENT{\textit{Define a Array of X}}
\STATE $CheckSet[SIZE\_OF\_SET + 1]$       \hfill \COMMENT{\textit{The array save the check set such as $\{1,3,7,8\}$}}
\newline
\FOR{$i  = 0$ to $SIZE\_OF\_PILE + 1$}
\STATE $X[i] \leftarrow{} 0$
\ENDFOR
\newline
\FOR{$i  = 0$ to $SIZE\_OF\_PILE + 1$}
\IF{$X[i] == 0$}
\FOR{$j = 0$ to $j < SIZE\_OF\_SET + 1$}
\STATE $X[i + CheckSet[j]] \leftarrow{} 1$
\ENDFOR
\ENDIF
\ENDFOR
\end{algorithmic}
\end{algorithm}

\subsection*{A simple proof}
It is simple to vindicate the PTFN algorithm. This paper can make sure its correctness ,only need to satisfy the conditions of the $P$-position's and $N$-position's definition(define 1).
\begin{itemize}
\item[] (1) In the subtraction games , $X[0]$ must be a terminal position ($P$-position), our deduction begin at this position .
\item[] (2) In PTFN algorithm , it gets $N$-position by one step which moved from $P$-position.
\item[] (3) PTFN means use $P$-position to find $N$-position, so from every P-position,every step move to N-position.
\end{itemize}
Through the above analysis, this paper vindicated the PTFN algorithm’s correctness.

\textbf{Example}
Next, using the method of PTFN analyzes \emph{Example 1} and tests the algorithm (in Figure.2).

\begin{figure}[htb]
\centering
\includegraphics[width=3.8in]{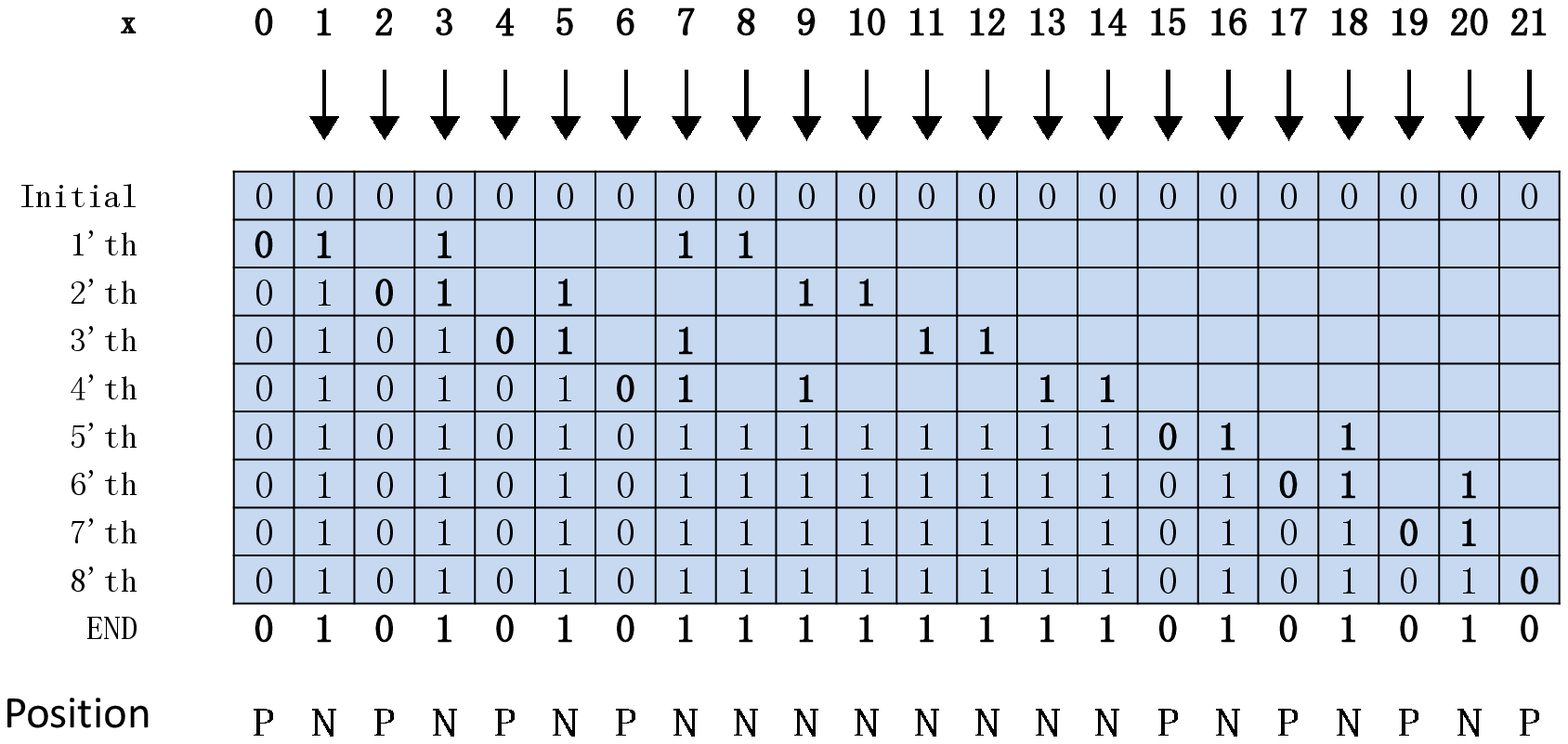}
\caption{The New Algorithm of the way with $S = \{1,3,7,8\}$}
\end{figure}

From the results obtained so far. The results of PTFN algorithm and Sprague-Grundy function are same.

\section{Some Application}
For one pile of the subtraction games, the PTFN algorithm is available. On the more than one pile subtraction games, the algorithm can be used, such as more than one pile subtraction games, Whyyof's Game and \emph{mis\'{e}re play rule} subtraction games.

\textbf{Sum of subtraction games},
first this paper give a formal description of a sum of games and then show how the PTFN algorithm for the component games.

Given more than one pile subtraction games, one can form a new game played according to the following rules.
\begin{itemize}
\item A given initial position is set up in each of the games($S_1,S_2,S_3,...$).
\item Players alternate moves.
\item A move for a player consists in selecting any one of the games and making a legal move in that game, leaving all other games untouched. Play continues until all of the games have reached a terminal position, when no more moves are possible.
\item The player who made the last move is the winner.
\end{itemize}

As an example, this paper used PTFN algorithm for the 2-pile game of subtraction games.
it only changes 1-dimension array $X[SIZE\_OF\_PILE]$ to 2-dimension array $X[SIZE\_OF\_PILE\_A][SIZE\_OF\_PILE\_B]$ , and row array presents one game ,raw array presents the other game. The Algorithm 2 shows main pseudocode.

\begin{algorithm}
\centering
\caption{PTFN for two piles}
\label{alg1}
\begin{algorithmic}[1]
\FOR{$i  = 0$ to $SIZE\_OF\_PILE\_A + 1$}
\FOR{$j = 0$ to $SIZE\_OF\_PILE\_B + 1$}
\IF{$X[i][j] == 0$}
\FOR{$k = 0$ to $k < SIZE\_OF\_SET\_A + 1$}
\STATE $X[i + CheckSetA[k]][j] \leftarrow{} 1$
\ENDFOR
\FOR{$t = 0$ to $t < SIZE\_OF\_SET\_B + 1$}
\STATE $X[i][j + CheckSetB[t]] \leftarrow{} 1$
\ENDFOR
\ENDIF
\ENDFOR
\ENDFOR
\end{algorithmic}
\end{algorithm}
Consider the sum of two subtraction games. One is $S = \{1,3,7,8\}$ (Example 1) and the pile has 15 chips. The other is $S = \{1,2,3,4\}$ and the pile has 15 chips. Thus,it obtains Figure.3 from different methods which PTFN algorithm and \emph{Sprague-Grundy theory}.
\begin{figure}[htb]
\centering
\includegraphics[width=3.8in]{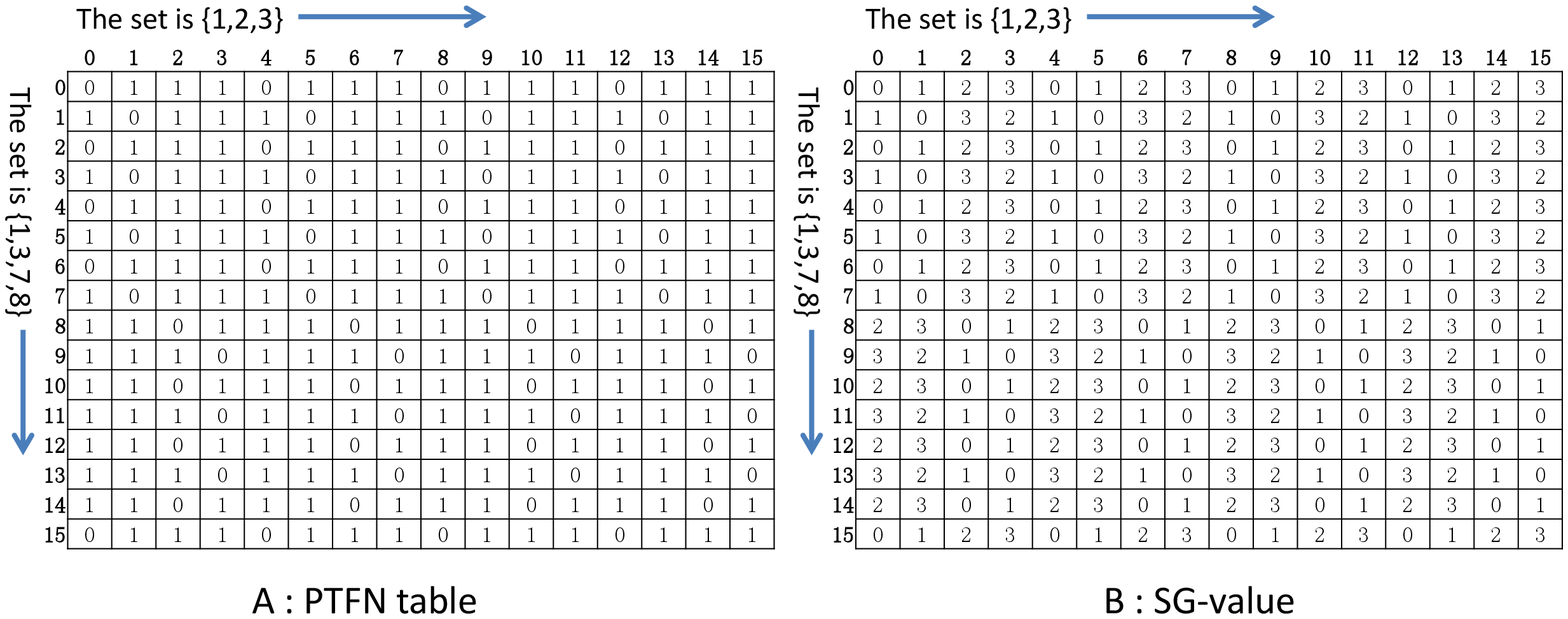}
\caption{The New Algorithm of the way}
\end{figure}

The Figure 3 results show that it is correct to use PTFN algorithm in the 2-pile game.

\textbf{Wythoff’s Game} introduced and solved by Willem A.Wythoff \cite{WA1} in 1907. And we can solve the game in the same way which solved 2-pile of subtraction games. And the only change is that they may take an equal number from both.

\textbf{Mis\'{e}re play rule for subtraction games}
For the games, if substraction set is $S = \{s_1,s_2,...,s_k\}$, we only set is $x[0]$ to $X[s_1]$ as $N$-position when the initialization of the PTFN algorithm.

\section{Epilogue}
This paper used a new algorithm to analysed the subtraction games. By comparison , this paper knows that for sums of subtraction games , Sprague-Grundy theory is better than PTFN algorithm, but for one pile subtraction games, PTFN has certain advantages over Sprague-Grundy theory.

In fact, $P$-position to find $N$-position is a good idea in many of games. we will use the PTFN algorithm to study these games in the future.

\providecommand{\href}[2]{#2}


\begin{thebibliography}{10}

\bibitem{RJN}
R.J.Nowakowski, \emph{History of Combinatorial Game Theory},Proceedings of the Board Game Studies Colloquium XI, Lisbon, 2008.

\bibitem{TF1}
Thomas S.Ferguson,\emph{Impartial Combinatorial Games}, Game Theory, Part $I$. Class notes for Math 167,(Fall 2000) 3-18.
%
\bibitem{EZ1}
E.Zermelo,\emph{\`{U}ber eine anwendung der mengenlehre auf die theorie des schachspiels},In Proc. 5th Int. Cong. Math. Cambridge 1912, Volume 2, .Cambridge University Press,(1913) 501-504.

\bibitem{EJR}
E.R.Berlekamp and J.H.Conway and R.K.Guy,\emph{winning ways for your mathematical plays},Volume.1 second ed..,A K Peters Ltd.2001

\bibitem{WA1}
Willem A. Wythoff,\emph{A modification of the game of Nim},Nieuw Arch. Wisk,Volume 7(1907) 199-202

\end{thebibliography}
     \end{document}